\documentclass{article}
\usepackage[utf8]{inputenc}
\usepackage{graphicx}
\usepackage{amssymb}
\usepackage{authblk}
\usepackage{textcomp}
\usepackage{gensymb}
\usepackage{tikz}
\usepackage{caption}
\usepackage{subcaption}
\usepackage[normalem]{ulem}
\usepackage{hyperref}
\hypersetup{
    colorlinks=true,
    linkcolor=blue,
    filecolor=magenta,  urlcolor=blue,
}

\title{Using homemade spectrometers to perform accurate measurements of discrete and continuous spectra}
\author[1]{Ana R. Romero Castellanos}
\author[2]{H. E. Castellanos}
\author[2]{C.E. Alvarez-Salazar}

\affil[1]{Universidad ECCI, Carrera 19 No. 49-20, 111311, Bogotá D.C, Colombia}
\affil[2]{Laboratorio de Física, Universidad Manuela Beltrán, Avenida Circunvalar No. 60-00, 110231, Bogotá D.C, Colombia}

\date{\today}
\newcommand*{\addheight}[2][.5ex]{%
  \raisebox{0pt}[\dimexpr\height+(#1)\relax]{#2}%
}
\begin{document}

\maketitle
\begin{abstract}
Homemade spectrometers are commonly used tools to analyze light sources and determine its physical characteristics. We perform an assessment of homemade spectrometers in terms of spectral resolution and accuracy in the determination of intensity, through the comparison of results with a spectroscope commonly used in the physics and chemistry labs. We found that the homemade spectrometer used is sufficiently accurate in wavelength, and can be used by undergraduate students to perform precise measurements as, for example, the spectrum of the Sun, leading to the determination of its temperature.

\end{abstract}

\section{Introduction}
Spectrometry, originally thought as the analysis of the interaction of radiation with matter with changing wavelength, but understood nowadays as the determination of the characteristics of any quantity as a function of wavelength or frequency, is one of the most relevant techniques to characterize a system.

Spectroscopy has found application in many fields, due to the fact that the properties of light emitted by a source can give quantitative information about a system. In this way, spectroscopy has been used for the study of atomic structure through the determination of emission and absorption spectra, the  analysis of chemical structures, the investigation of chemical equilibrium and the kinetics of chemical reactions~\cite{robert1961fundamentals,perkampus2013uv}. Spectroscopy has found applications in environmental analysis to determine the acidity of water or the presence of metallic compounds in a sample~\cite{mesilaakso1997application,li2016applications}, in medicine has been used for diagnostic and therapeutic applications as well as for pharmaceutical analysis~\cite{matousek2013recent,kim2020optical}, and in astronomy has been used, both long ago and nowadays, to determine the chemical composition of stars or planets~\cite{gonzalez2006chemical,alibert2017formation,mishenina2021chemical}.

In this way, spectroscopy is a very important tool for different professionals and scientists, and this makes teaching the basics of spectroscopy a must in physics courses. Furthermore, spectroscopes are broadly used in the modern physics lab to understand the structure of the atom and introducing students to the concepts of quantum mechanics and the quantization of light~\cite{galindo2012quantum,scholz2020classical}.

Nevertheless, a precision spectrometer is an expensive device which many institutions can not afford, and this situation has raised different experimental designs to supply this need. On the other hand, it could be very enlightening for students, as mentioned in~\cite{taha2017simple, balado2019spectrophotometers}, to build a spectrometer to understand the measurement process and the physics behind the device used to characterize a given chemical solution, as performed by the authors of the same references. 

Taking these considerations into account, homemade spectrometers have been designed and built to perform different experiments. A measurement of wavelengths of Balmer series for hydrogen has been obtained (with very low relative errors) using spectrometers with transmission/reflection diffraction gratings~\cite{onorato2015measuring}. A determination of the concentration of ferrous ($\textrm{Fe}^{2+}$) and sodium ($\textrm{Na}^{+}$) ions in saline solutions and natural water samples have been made in
~\cite{de2017handheld}, using a low cost spectroscope connected to a smartphone. On the same vain, Hosker~\cite{hosker2018demonstrating} has built a shoebox spectrophotometer using the light sensor on a smartphone to perform measurements of absorbance as a function of concentration for different solutions. 

Certainly, with the COVID-19 pandemic, virtual labs using simulators, homemade experiments and the use of both smartphone and online apps have become common tools in education to perform all kind of experiments in physics~\cite{rosi2020video,iqbal2020teaching,tan2021bringing,klein2021studying,shidiq2020simple,koohkan2020fabrication,bastidas2020construccion}, chemistry~\cite{andrews2020experimenting,sunasee2020challenges}, microbiology~\cite{lopez2021home,koort2021redesigning}, acoustics~\cite{korman2021teaching}, and all fields of knowledge. All these strategies have been assessed in terms of their effectiveness and usefulness for scientific education~\cite{gordy2021science,tanik2020teaching}. Nevertheless, it would be good to ask how different are the results obtained with a homemade measurement device to the ones measured using specialized equipment commonly used in the undergraduate physics and chemistry labs, and if the results obtained can be used by students to perform accurate measurements.

In this work, we present an assessment of the results obtained with a homemade spectrometer
and make a comparison with the results taken using a common physics lab device. We have captured the spectrum of three different light sources (hydrogen, helium and neon) using the Public Lab foldable mini-spectrometer \cite{PublicLab} available at \hyperlink{publiclab.org/wiki/foldable-spec}{publiclab.org/wiki/foldable-spec} and using the do-it-yourself (DIY) material analysis tool Spectral Workbench by Public Lab, available at \hyperlink{spectralworkbench.org/}{spectralworkbench.org/}, and compare the results with the ones obtained with the Pasco wireless spectrometer, using the free Pasco Spectrometry Software available at \hyperlink{pasco.com/products/software/spectrometry}{pasco.com/products/software/spectrometry}. After comparison, we conclude that the results obtained with the homemade spectrometer are sufficiently accurate and can be used by undergraduate students to perform more complex measurements for both discrete and continuous light sources.

In this sense, we propose to use the homemade spectrometer to determine the spectrum of sunlight and determine the value of the temperature of the Sun, a measurement which could be of interest both for physics and astronomy undergraduate students. We suggest a comparison of the results with the Standard Solar Spectra, freely available at \hyperlink{pveducation.org/pvcdrom/appendices/standard-solar-spectra}{pveducation.org/pvcdrom/appendices/standard-solar-spectra}, in order to compare the fits obtained by a homemade experiment and a more sophisticated one, what reinforces the skills of data analysis and evaluation.

This experiment and its related data analysis can help students to understand the phenomena of light diffraction and its use for the measurement of light properties, the characteristics and differences of discrete and continuous spectra, the atomic structure and the characteristics of black-body radiation.

This paper is organized as follows. In Sec.~\ref{sec:Experimental} we present the
experimental setup used, describe the homemade spectrometer and the measurement process used, and make a slight description of the capture of spectra using the wireless Pasco spectrometer. In Sec. \ref{sec:Comparison} we present the results obtained using both the Pasco Wireless Spectrometer and the homemade spectrometer for three different light sources, and compare the results in order to assess the viability of using the homemade spectrometer to perform more complex studies. In section~\ref{sec:SunSpectrum} we show the results for the measurement of the spectrum of sunlight, from which a blackbody spectrum has fitted in order to determine the surface temperature of the Sun. 
Finally, we present our conclusions in Sec.~\ref{sec:conclusions}.

\section{Experimental setup and calibration}\label{sec:Experimental}

The first step to develop this home experiment to analyze light sources is to build the measuring device using the foldable mini-spectrometer designed by Public Lab and available at \hyperlink{https://publiclab.org/sites/default/files/8.5x11mini-spec3.8.pdf}{https://publiclab.org/sites/default/files/8.5x11mini-spec3.8.pdf}. In order to build the spectrometer, students need black cardboard, a DVD-R, scissors and glue or adhesive tape, which makes the instrument inexpensive, easy to build and suitable for use by any student of any institution.

For our homemade experiment we have chosen to use a pre-designed spectrometer already available and not to design a different configuration, as done, for example, in~\cite{taha2017simple,balado2019spectrophotometers,ju2020fabrication,rosi2016we}, in order to perform an assessment of the results obtained with the homemade apparatus, and compare these measurements with the ones obtained with a commonly used equipment in physics teaching lab, the PASCO wireless spectrometer.

Once the spectrometer has been built, the online application \hyperlink{spectralworkbench.org/}{spectralworkbench.org} should be used to get the spectrum of a light source. Using a laptop webcam or a mobile phone camera, it is possible to get the spectrum (a curve of light intensity as a function of wavelength) for different light sources. The calibration process is fundamental, as the image pixels are converted in nanometers, giving results of acceptable precision, as analyzed later in this paper. Students can perform the calibration at home by direct comparison of the light spectrum emitted by a compact fluorescent lamp (CFL) with the mercury reference spectrum used in the app, matching the main blue (436 nm) and green (546 nm) lines emitted by the mercury gas contained in the CFL with the reference spectrum given in the web-based application. 

\begin{figure}[htp]
\centering
\includegraphics[width=8cm]{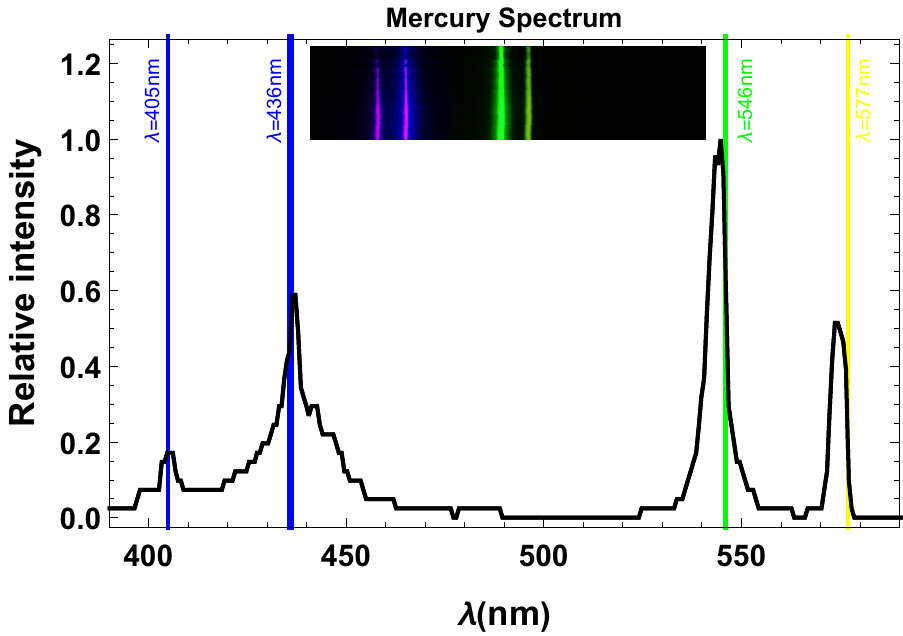}
    \caption{Mercury spectrum obtained using the homemade spectrometer and the web application \href{spectralworkbench.org}{spectralworkbench.org}. The vertical colored lines show the main emission lines for this element in the visible range, where the corresponding wavelengths are indicated. The spectrum obtained in the application is shown in the insert.}
    \label{fig:MercuryCasero}
\end{figure}

In our experiment, we have used a mercury spectral tube to perform the calibration, finding the results in  figure~\ref{fig:MercuryCasero}, where a plot of relative intensity as a function of wavelength is shown, and we can compare the wavelengths of the peaks with the spectrum presented in \hyperlink{https://physics.nist.gov/PhysRefData/Handbook/Tables/mercurytable2.htm}{Basic atomic spectroscopy data}. In this graph, we can see the emission spectrum as a continuous black line showing four different peaks, matching very closely the reference emission lines at 405, 436, 546 and 577 nm, approximately, indicated by vertical colored lines in the graph. Additionally, we have inserted the line spectrum, as obtained with 
\hyperlink{spectralworkbench.org/}{spectralworkbench.org} using a webcam, where we can see the strong green and blue lines used to perform the calibration.

From figure~\ref{fig:MercuryCasero} we can see that the mercury spectrum, as obtained through the homemade spectrometer and the web-based application \hyperlink{spectralworkbench.org}{spectralworkbench.org}, is composed by a set of four intensity peaks at 403, 438, 545 and 575 nm with a specific width, characterizing the resolution of the spectrometer, a property which we discuss later in this paper. We can see that these results have a maximum percentage error of 1.2\%, showing a very good approximation to the position of peaks reported elsewhere~\cite{lide2004crc}, what makes the homemade spectrometer a good option to characterize other light sources.

\section{Characterization of discrete sources using the homemade spectrometer}\label{sec:Comparison}

In this section, we present the results obtained with the homemade spectrometer for the analysis of light emitted by three different spectral tubes: hydrogen, helium, and neon. We also assess the characteristics of the homemade spectrometer and compare the results with the ones obtained using the PASCO wireless spectrometer, commonly used in physics and chemistry teaching labs.

\begin{figure}[htb!]
\centering
\includegraphics[width=8cm]{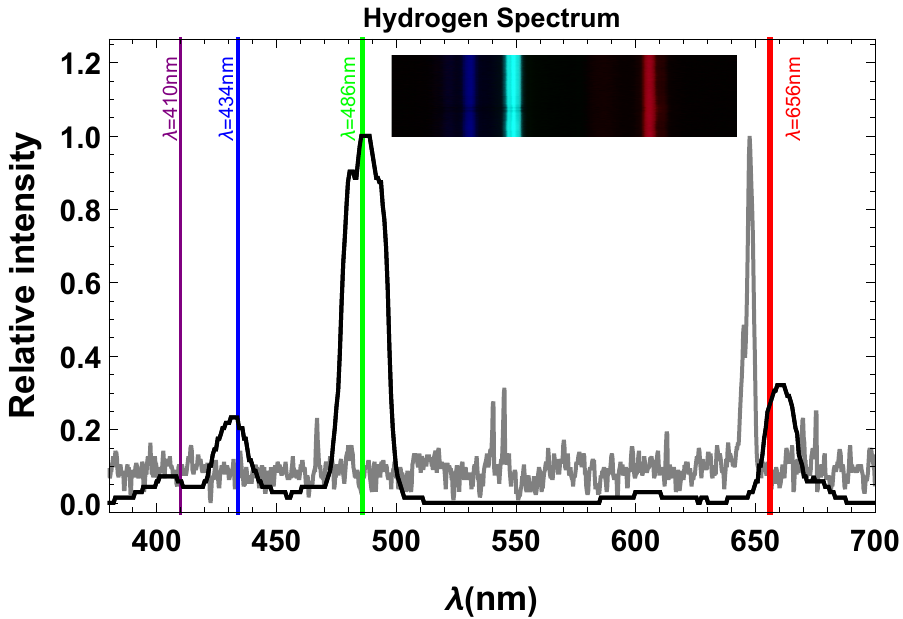}
    \caption{The hydrogen spectrum obtained with the homemade spectrometer (black) and with the PASCO spectrometer (gray). The vertical colored lines show the main emission wavelengths for this element, written alongside each line. The insert shows the capture in the web-base application \href{spectralworkbench.org}{spectralworkbench.org}.}
    \label{fig:HidrogenoCasero}
\end{figure}

In figure~\ref{fig:HidrogenoCasero} we can see the hydrogen spectrum obtained using both the homemade and PASCO spectrometers, in black and gray, respectively. In this graph, we can see the spectral lines reported by~\cite{lide2004crc} as vertical colored lines, where the corresponding wavelengths are indicated alongside each line. The line spectrum for this source is shown in the insert. From this graph, we can see that the spectrum obtained with the homemade spectrometer consists of four peaks with wavelengths 403, 428, 486 and 667 nm, which correspond to radiation emitted in the transition of electrons from a higher energy level to a lower energy level\footnote{The violet (403 nm) line corresponds to a transition from the 6\textsuperscript{th} to the 2\textsuperscript{nd} energy level, the blue line (428 nm) to a transition from the 5\textsuperscript{th} to the  2\textsuperscript{nd}, the green (486 nm) is produced in the 4\textsuperscript{th} to the 2\textsuperscript{nd} transition, and the red (667 nm) in the 3\textsuperscript{rd} to 2\textsuperscript{nd} energy level transition. This analysis has already done, for example, in~\cite{scholz2020classical,onorato2015measuring}.}. The lines obtained deviate from the most intense emission lines of hydrogen in, at most, 2\%. Again, this deviation is very small and reinforces the observation that the homemade spectrometer gives results of sufficient accuracy to perform the analysis of different light sources, which can reinforce the knowledge of modern physics in undergraduate students.

In this sense, we have found the resolving power of the homemade spectrometer, defined as
\begin{equation}\label{eq:resolving}
    R=\frac{\lambda}{\Delta\lambda},
\end{equation}
where $\Delta \lambda$ is the smallest resolvable wavelength difference between two monochromatic lines observable in the spectrum~\cite{robertson_2013}. To determine $\Delta \lambda$ for the homemade spectrometer, we found the full width at half maximum (FWHM) by fitting a Gaussian distribution to each peak and finding the position of the maximums, as can be seen in table~\ref{tab:AjusteHidrogeno}, where the best fit for each data set around the corresponding wavelength is shown as a continuous line, and the FWHM is indicated in each graph. From these results we have found a resolving power of 18 for a wavelength of 403 nm, 25 for 428 nm, 27 for 486 nm and 46 for the 667 nm wavelength. These results show that the greatest resolving power of the spectrometer corresponds to the red color which, nevertheless, has the lowest intensity in the spectrum. 

It is important to note that the results of the homemade spectroscope are accurate in terms of wavelength, but not in relative intensity, as can be seen in figure~\ref{fig:HidrogenoCasero}, which shows its maximum intensity at a wavelength of 486 nm, the green line of the spectrum. On the other hand, the PASCO spectroscope shows the most intense line in the red color, at a wavelength of 648 nm, very close to the value reported in~\cite{lide2004crc}. From this result, and the absence of the orange line in the mercury spectrum in figure~\ref{fig:MercuryCasero}, we conclude that there is a suppression of intensity in the measurement of higher wavelengths, a point which will be discussed later in this paper.

\noindent
\begin{table}[ht]
\centering
\begin{tabular}{|c|c| }
\hline
\small $\lambda_{\textrm{exp}}=403$ nm &  $\lambda_{\textrm{exp}}=428$ nm\\
      \addheight{\includegraphics[width=50mm]{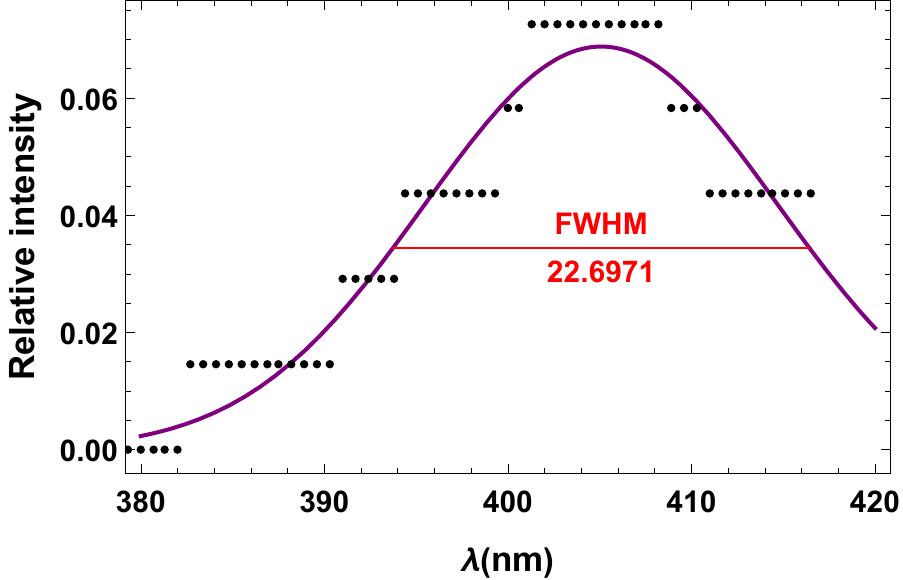}}
       &
      \addheight{\includegraphics[width=50mm]{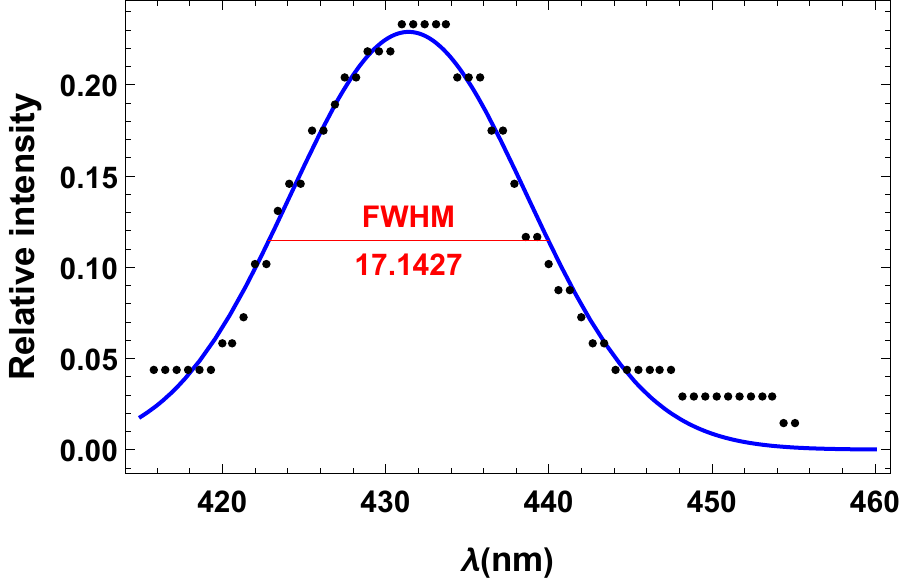}}
      \\
      \hline
      \small $\lambda_{\textrm{exp}}=486$ nm &  $\lambda_{\textrm{exp}}=667$ nm\\
      \addheight{\includegraphics[width=50mm]{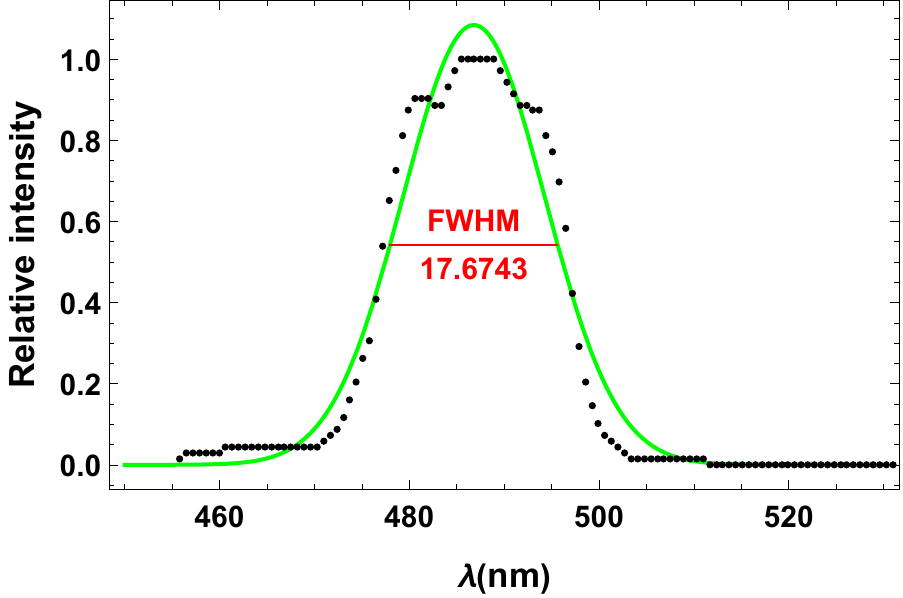}}
        &\addheight{\includegraphics[width=50mm]{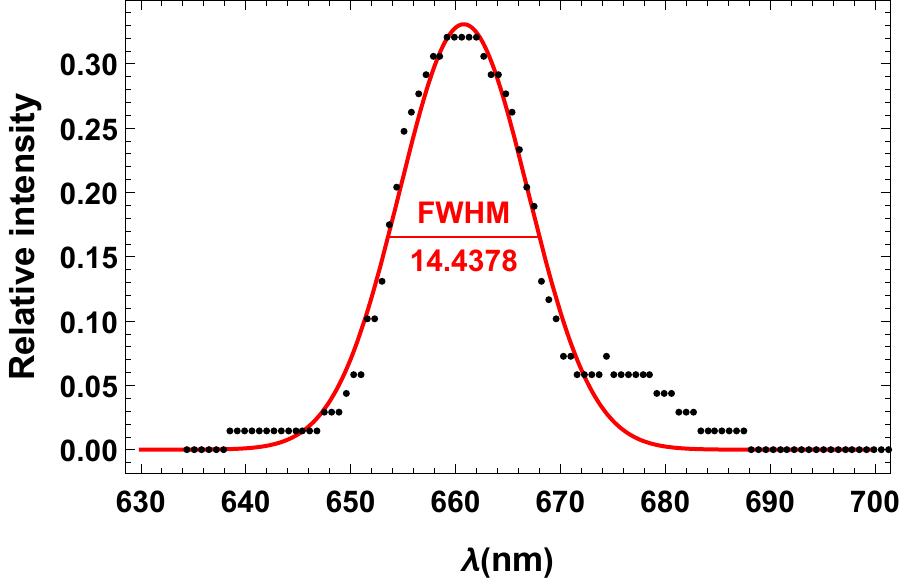}}\\
        \hline
\end{tabular}
\caption{Best fits for each of  hydrogen peaks observed from figure~\ref{fig:HidrogenoCasero}, where the resonance wavelength observed with the homemade spectrometer is shown in the upper part, and the FWHM for each peak is shown.}
\label{tab:AjusteHidrogeno}
\end{table}

\begin{figure}[htb!]
\centering
\includegraphics[scale=1.0]{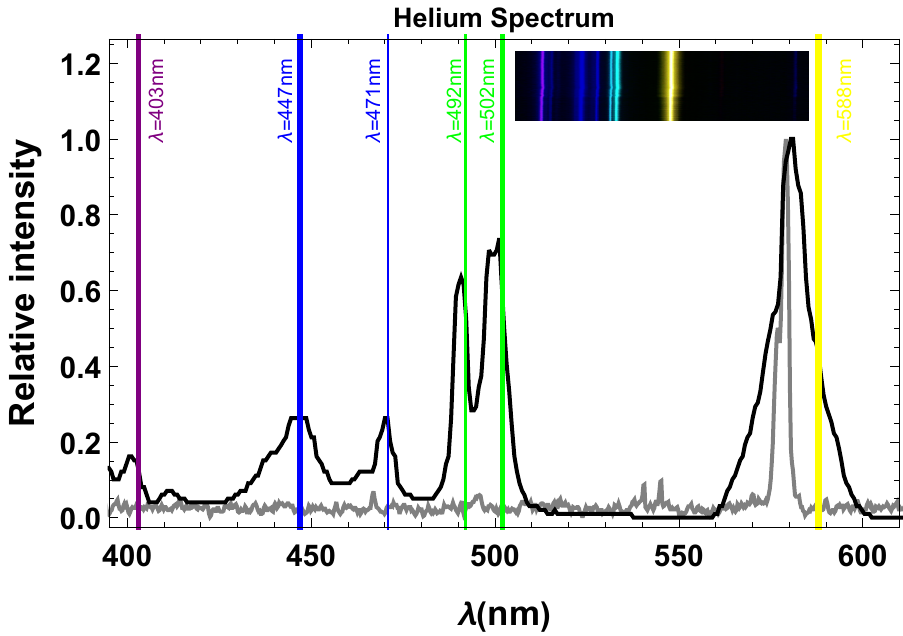}
    \caption{Helium spectrum obtained with the homemade spectrometer (black), and its comparison with the PASCO spectrometer (gray) and reference emission lines~\cite{lide2004crc}. The spectrum obtained is shown in the insert.}
    \label{fig:HeliumCasero}
\end{figure}

In figure \ref{fig:HeliumCasero} we show the spectrum obtained with the homemade spectrometer (black continuous line) for the helium spectral tube, and the corresponding comparison with the PASCO spectrometer (gray continuous line) with the emission lines reported in \cite{lide2004crc} indicated by vertical colored lines with the corresponding wavelengths written alongside. The line spectrum for this source is shown in the insert. From this graph we can see that the homemade spectrometer gives six peaks at  wavelengths 400, 446, 470, 490, 499 and 580 nm, which deviate at most 1.4\% with respect to the lines reported in~\cite{lide2004crc}.

From the helium spectrum shown in figure~\ref{fig:HeliumCasero} we can notice several characteristics. We can see interference effects in the green lines of the spectrum, separated approximately 10 nm (less than the FWHM of the spectrometer for the green lines, approximately 18 nm, as can be seen in table~\ref{tab:AjusteHidrogeno}). On the other hand, it is important to remember that the spectrum obtained with the homemade spectrometer is accurate in terms of wavelengths but not in intensities. This remark is revealed in the helium spectrum too, as the blue line with measured wavelength 446 nm should have, according to the measurements reported in~\cite{lide2004crc}, an intensity greater than the intensities of the two green lines at 492 nm and 502 nm, which are shown with higher intensity in the spectrum obtained with the homemade spectrometer.

This effect, and also the absence of the red lines at 667, 687 and 707 nm persistent in the helium spectrum shows the effects of the filter in the webcam or mobile phone camera. This is due to the fact that digital cameras used in laptops and smartphones have a decrease in their relative spectral sensitivity, describing their relative efficiency of light detection for different wavelengths, for wavelengths in the interval between 600 and 700 nanometers, corresponding to the red part of the visible spectrum. For this color, the relative sensitivity is at most 50\%, while the relative sensitivities for green and blue reach, approximately, 100\% and 85\%, respectively~\cite{jiang2013space,rowlands2020color}. This also explains why the web-based application \hyperlink{spectralworkbench.org}{spectralworkbench.org}, when used with a laptop or smartphone camera, shows the green peaks with greater intensity than the blue or the red peaks, even though their intensity should not be that high, as can be seen in figure~\ref{fig:HidrogenoCasero} with the red line measured at 667 nm (which should be the one with maximum intensity). 

\begin{figure}[htb!]
\includegraphics[width=12cm]{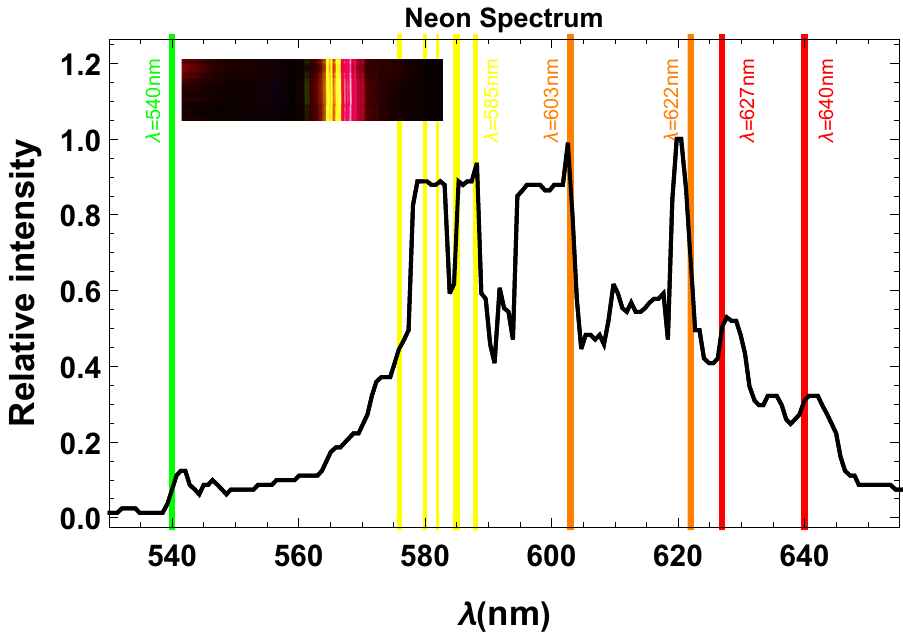}
    \caption{Neon spectrum obtained with the homemade spectrometer (black), and its comparison with the reference emission lines at~\cite{lide2004crc}. The insert shows the spectrum obtained.}
    \label{fig:NeonCasero}
\end{figure}

To close this section, we present in figure~\ref{fig:NeonCasero} the spectrum obtained for the neon spectral tube, drawn as a continuous black line. In the same graph, the reference spectrum is indicated by the vertical colored lines with the corresponding wavelengths indicated alongside, and the spectrum observed in the web-based application \hyperlink{spectralworkbench.org}{spectralworkbench.org} is shown in the insert. We have chosen to avoid the inclusion of the PASCO spectrum in this case in order to emphasize the characteristics of the neon spectrum, and perform an analysis with the same tools which a student performing the experiment would have. We can see that this spectrum, as measured by the homemade spectrometer, consists of seven intensity peaks with wavelengths 541, 581, 587, 600, 621, 629 and 642 nm, which deviate, at most, 1\% with the lines reported in reference~\cite{lide2004crc}.

The characteristics of the homemade spectrometer, obtained from the analysis of light emitted by the hydrogen and helium spectral tubes, are shown completely in the neon spectrum of figure~\ref{fig:NeonCasero}. 

The spectrometer shows an interference between sets of spectral lines which are very close, as the yellow lines shown in the spectrum. The first observed peak seen in the yellow part of the visible spectrum corresponds to the interference of three spectral lines at wavelengths 576, 580 and 582 nm, and the second observed peak in the same color corresponds to the interference of radiation emitted at wavelengths of 585 and 588 nm, as reported in~\cite{lide2004crc}. It is important to notice that, even though we can distinguish two peaks, they are not fully disentangled, as the five wavelengths in the yellow part of the visible spectrum are very close, and the resolving power of the homemade spectrometer is not enough to identify each line independently. This behavior is also seen in the orange line measured at 600 nm with the homemade spectrometer, which corresponds to the interference of emission lines at 597 and 598 nm, very close to the observed wavelength, explaining the wide peak observed around 600 nm. 

Another characteristic previously discussed is the effect of the filter in the webcam or mobile phone camera, which makes the spectrum sufficiently accurate in wavelength, but not in terms of intensity. This aspect is revealed in the small intensities shown in the red part of the spectrum (emission lines at 627 and 640 nm), which are shown very low in comparison with the yellow and orange peaks, even though the red line near 627 nm should have the same intensity as the two orange lines, as reported in \hyperlink{https://physics.nist.gov/PhysRefData/Handbook/Tables/mercurytable2.htm}{Basic atomic spectroscopy data}. 

An aspect which calls for attention in the results obtained with the homemade spectrometer is the attenuation in intensity observed for the green peak near 540 nm and the red peak near 640 nm. According to experimental data reported in~\cite{lide2004crc}, these two lines, together with the yellow line near 585 nm, should be the more intense lines and have equal intensities. The spectrum in figure~\ref{fig:NeonCasero} shows a strong attenuation of these two lines, but the reason for each one is very different. For the green line near 540 nm, the attenuation is due to the position and geometry of the homemade spectrometer, which bother the observation of this line. For the case of the red line near 640 nm, we observe a strong attenuation due to the filter effects of the webcam or mobile phone camera.

It should be desirable for students to perform a correction of the intensities measured, according to the methods explained, for example, in~\cite{jiang2013space, rowlands2020color}, in order to improve their experimental data. 

\section{Sunlight spectrum and Blackbody radiation}\label{sec:SunSpectrum}
The spectra analyzed so far correspond to the intensity as a function of the wavelength of light emitted by atomic sources, and are composed of different discrete lines emitted when an electron makes a transition from a higher energy level to a lower energy level. In this section,
we use the homemade spectrometer to analyze a light source with a continuous spectrum, the Sun, and determine the characteristics of its radiation.

\begin{figure}[bht!]
\includegraphics[width=12cm]{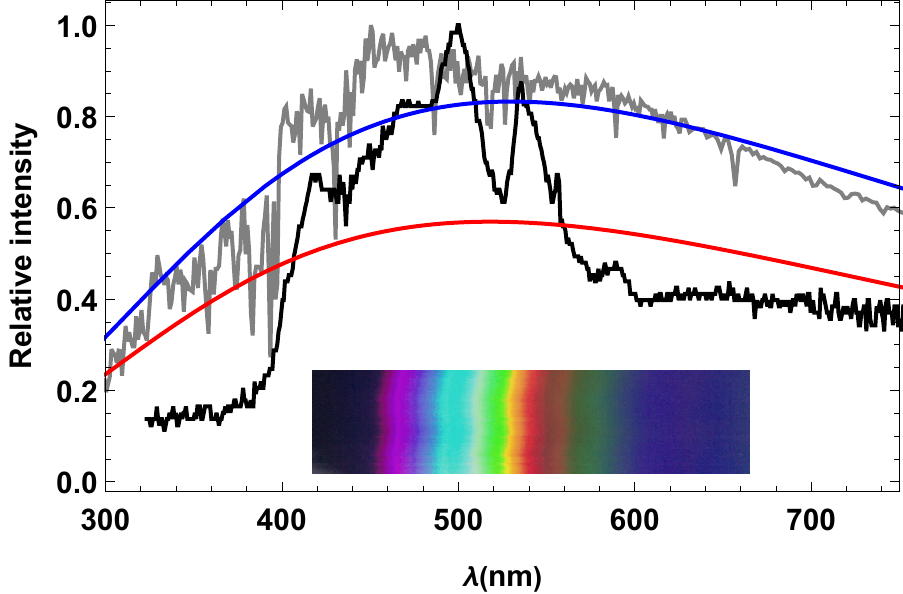}
    \caption{Sunlight spectrum obtained with the homemade spectrometer (black), and its comparison with the standard solar spectrum (gray). The red and blue lines show the blackbody radiation spectrum for the homemade and solar spectra, respectively.}
    \label{fig:SolarComparado}
\end{figure}

In figure~\ref{fig:SolarComparado} we show the sunlight spectrum obtained with the homemade spectrometer (continuous black line) and the blackbody radiation spectrum obtained for this data set, as a continuous red line. In the same graph, we can see the standard solar spectrum~\cite{honsberg2021standard} as a continuous gray line and its corresponding fit using the blackbody radiation spectrum, as a continuous blue line.

In this figure, we can see that the measurement taken with the homemade spectrometer for the spectrum of sunlight, shown as a continuous black line in the graph, gives an appropriate characterization of this light source for wavelengths between 400 and 540 nm, where the attenuation of relative intensity is not very significant and the observations with the homemade spectrometer are very close to the standard solar spectrum. As was previously mentioned in the analysis of sources with discrete spectrum, the attenuation observed for the red part of the visible spectrum is due to the effect of camera filters. This strong attenuation causes a drop in the red part of the spectrum which affects the fit of the data set, giving a smaller maximum than the observed from the raw data set.

From the fit of the blackbody radiation spectrum performed on the data set obtained with the homemade spectrometer (red line) and its comparison with the corresponding fit to the standard solar spectrum (blue line), we can see, besides the strong attenuation for higher wavelengths, a displacement of the maximum of the graph, and this will give a lower value of the temperature of the Sun. More specifically, from the fit to the blackbody radiation spectrum to the data obtained with the homemade spectrometer, we obtained a Sun temperature of $(5591\pm 70)$ K, with a percentage error of approximately 3\% with respect to the reported temperature of the Sun~\cite{kuhn1988surface,holmberg2006colours}.  

\section{Conclusions}\label{sec:conclusions}

In this work we have shown that homemade spectrometers can be used to perform an accurate characterization of both discrete and continuous light sources. We have seen that the homemade spectrometer is accurate in terms of wavelength, but presents an attenuation in intensity, specially for the red part of the visible spectrum, due to the effect of laptop or mobile phone cameras used by students performing the experiment

We have seen that the spectral resolution is enough to disentangle discrete lines with separation greater than 10 nm, but interference effects for lines with separation very close to this threshold often appear. This characteristic allows students to understand the physical processes behind the spectroscope, and the phenomenon of interference between light waves. 

Additionally, we have seen that filters in laptop and mobile phone cameras cause major effects in the properties of spectra, especially in the red part of the spectrum, where attenuation can decrease the intensity of light up to 50\%.

Finally, we have seen that students can perform the characterization of continuous light sources as the Sun, and use the measured spectra to understand the properties of blackbody spectrum and determine the surface temperature of the Sun.

\section*{Acknowledgments}
A.R.R.C is grateful for the academic support of the Basic Sciences Department at Universidad ECCI. H.E.C and C.E.A.S are grateful for the support of Laboratory coordination at Universidad Manuela Beltrán.
\bibliographystyle{unsrt}
\bibliography{References.bib}

\end{document}